\documentclass[epj]{svjour}
\usepackage{amsmath}
\usepackage{amssymb}
\usepackage{graphicx}
\allowdisplaybreaks[4]

\usepackage{bm}

\newcommand{\dn}{{d_N}}
\newcommand{\ds}{{d_S}}

\newcommand{\kfs}{{k_{FS}}}
\newcommand{\Zb}{{Z}}

\newcommand{\knp}{{k^+_N}}
\newcommand{\knm}{{k^-_N}}

\newcommand{\ksp}{{k^+_S}}
\newcommand{\ksm}{{k^-_S}}

\newcommand{\ub}{{\bar u}}
\newcommand{\vb}{{\bar v}}

\newcommand{\mr}{m_r}

\newcommand{\npa}{{\knp}}
\newcommand{\nma}{{\knm}}

\newcommand{\spa}{{\ksp}}
\newcommand{\sma}{{\ksm}}

\newcommand{\cnp}{{c^+_N}}
\newcommand{\snp}{{s^+_N}}

\newcommand{\csp}{{c^+_S}}
\newcommand{\ssp}{{s^+_S}}

\newcommand{\cnm}{{c^-_N}}
\newcommand{\snm}{{s^-_N}}

\newcommand{\csm}{{c^-_S}}
\newcommand{\ssm}{{s^-_S}}

\newcommand{\atz}{{\Zb\kfs}}

\newcommand{\MyTextIt}{}

\begin{document}

\title{Quasiparticle states in superconducting superlattices}%

\author{Mihajlo Vanevi\' c%
\thanks{\emph{Present address:} Department of Physics, University of
    Basel, Klingelbergstrasse 82, 4056 Basel, Switzerland;
    \email{mihajlo.vanevic@unibas.ch}}%
\and Zoran Radovi\' c
} % end \author
%
%\email{mihajlo.vanevic@unibas.ch}
\institute{Department of Physics, University of Belgrade, P. O. Box 368,
11001 Belgrade, Serbia and Montenegro}

\date{Received: \today / Revised version: date}

\abstract{%
The energy bands and the global density of states are computed for
superconductor / normal-metal superlattices in the clean limit.
Dispersion relations are derived for the general case of insulating
interfaces, including the mismatch of Fermi velocities and effective
band masses. We focus on the influence of finite interface
transparency and compare our results with those for transparent
superlattices and trilayers. Analogously to the rapid
\MyTextIt{variation on the atomic scale} of the energy dispersion with
layer thicknesses in transparent superlattices, we find strong
oscillations of the almost flat energy bands (transmission
resonances) in the case of finite transparency. In small-period
transparent superlattices the BCS coherence peak disappears and a
similar subgap peak is formed due to the Andreev process. With
decreasing interface transparency the characteristic double peak
structure in the global density of states develops towards a gapless
BCS-like result in the tunnel limit. This effect can be
used as a reliable STM probe for interface transparency.%
 % end of PACS codes
}% abstract

%\keywords{superconducting superlattice interface transparency band
% structure density of states}

\maketitle

%%%%%%%%%%%%%%%%%%%%%%%%%%%%%%%%%%%%%%%%%%%%%%%%%%%%%%%%%%%%%%%%%%%%%%%%%%%%%
%%%%%%%%%%%%%%%%%%%%%%%%%%%%%%%%%%%%%%%%%%%%%%%%%%%%%%%%%%%%%%%%%%%%%%%%%%%%%
\section{\label{sec:Intro}Introduction}

The artificial S/N superlattices consisting of alternating
superconductor (S) and normal-metal or semiconductor (N) layers have
been studied for some time already
\cite{art:BaranovBuzdin,art:TakahashiTakahashi1,art:RadovicLedvij,art:Bulaevski90,art:Buzdin92,art:TanakaTsukada,art:Stojkovic,art:Kummel94,art:Kuplevakshky},
see also \cite{art:Koperdraad,art:KettersonBook}. The recent
advancement of nanofabrication technology and experimental
techniques \cite{art:Moussy02}, as well as intrinsically layered
structure of high-$T_c$ superconductors
\cite{art:Huang03,art:Buschmann92,art:Kashiwaya} has reinvigorated
the long standing interest in understanding the effects inherent to
clean superconducting heterostructures
\cite{art:Mortensen,art:Gyorffy96,art:GunsenhKummel94}. The size and
coherence effects have been studied recently for double barrier SNS
and NSN junctions in the clean limit based on the solutions of
Gor'kov and Bogoliubov--de Gennes (BdG) equations
\cite{art:TanakaTsukada1,art:Brinkman,art:RadovicGrcka,art:BozovicRadovic,art:BozovicRadovicNew,art:RadovicPhysC}.

In this paper we extend the previous approach of Tan\-aka and
Tsukada \cite{art:TanakaTsukada} and Plehn {et al.}
\cite{art:Kummel94} to the more general case of superlattices with
finite interlayer transparency. We present comprehensive and
systematic analysis of the influence of interface transparency on
the quasiparticle band structure and density of states for wide
range of the superlattice parameters. Due to the phase coherence of
electronic wave functions  the energy spectrum is gapless in
superlattices with thin S layers and transparent interfaces
\cite{art:RadovicPhysC}, and splits into almost flat bands
(transmission resonances) with decreasing transparency. For thick S
layers, the subgap bands are formed due to the Andreev reflection
\cite{art:Andreev} which leads to \MyTextIt{the conversion of  Cooper
pairs} in superconducting layers into correlated electrons and holes
in the normal layers. \MyTextIt{Whereas the calculations are performed
in the clean limit, the influence of impurities on the density of
states} can be taken into account by replacing the superconducting
coherence length with an effective one, as shown by Halterman and
Valls \cite{art:Halterman} in comparison with experiments of Moussy
{et al.} \cite{art:Moussy02}. Our results for density of states in
superlattices with layer thicknesses smaller than the
superconducting correlation length, qualitatively confirm main
features previously obtained by Bulaevskii and Zyskin
\cite{art:Bulaevski90} and Buzdin {et al.} \cite{art:Buzdin92} for
atomic-scale layered systems within the tight binding approximation.

%%%%%%%%%%%%%%%%%%%%%%%%%%%%%%%%%%%%%%%%%%%%%%%%%%%%%%%%%%%%%%%%%%%%%%%%%%%%%
%%%%%%%%%%%%%%%%%%%%%%%%%%%%%%%%%%%%%%%%%%%%%%%%%%%%%%%%%%%%%%%%%%%%%%%%%%%%%
\vspace*{-3mm}
\section{\label{sec:Model}The model}

The system under consideration is an S/N superlattice in the clean
limit, consisting of alternating superconducting and normal-metal
(or semiconductor) layers of thickness $d_S$ and $d_N$, with
insulating interfaces modelled as thin potential-energy barriers.
The superconducting layers are characterized by constant pair
potential $\Delta_0$, and zero phase difference, $\phi=0$, is
assumed across the superlattice. Effective band masses and
electrostatic potentials of the two metals are $m_S$ ($m_N$) and
$U_S$ ($U_N$), respectively. The superlattice is uniform in the $x-y$
plane and the $z$ axis is perpendicular to the layers.

Quasiparticle propagation in the superlattice is described by the
Bogoliubov--de Gennes equation
%%%%%%%%%%%%%%%%%%%%%%%%%%%%%%%%%%%%%%%%%%%%%%%%%%%%%%%%%%%%%%%%%%%%%%%%%%%%%
\begin{equation}
\begin{pmatrix}
    H_0({\bf r}) & \Delta({\bf r}) \\
    \Delta^*({\bf r}) & -H_0^* ({\bf r})
\end{pmatrix}
\Psi({\bf r}) = E \Psi({\bf r}),
\end{equation}
%%%%%%%%%%%%%%%%%%%%%%%%%%%%%%%%%%%%%%%%%%%%%%%%%%%%%%%%%%%%%%%%%%%%%%%%%%%%%
where $\Psi({\bf r})=\big(u({\bf r}), v({\bf r})\big){}^T$ is the
two-component wave function in the \MyTextIt{electron-hole space}, the
quasiparticle energy $E$ is measured with respect to the chemical
potential $\mu$, and the hamiltonian $H_0$ within the superlattice
period $a=d_N+d_S$, for $z\in(-d_N,d_S)$, is given by
%%%%%%%%%%%%%%%%%%%%%%%%%%%%%%%%%%%%%%%%%%%%%%%%%%%%%%%%%%%%%%%%%%%%%%%%%%%%%
%\begin{equation*}
\begin{multline}
H_0({\bf r}) = - \nabla \frac{\hbar^2}{2m({\bf r})} \nabla
+ \hat W \, [ \delta(z) + \delta(z+d_N) ]
\\
+U({\bf r})- \mu.
\end{multline}
%\end{equation*}
%%%%%%%%%%%%%%%%%%%%%%%%%%%%%%%%%%%%%%%%%%%%%%%%%%%%%%%%%%%%%%%%%%%%%%%%%%%%%
The first term is the quasiparticle kinetic energy in the
effective mass approximation \cite{art:Mortensen,art:effMass}, the
second term, with $\hat W=\hbar^2 k_{FS} Z / 2m_S$, describes
finite transparency of S-N interfaces modelled as
$\delta$-function potential barriers, and dimensionless parameter
$Z$ measures the barrier strength. Fermi energies in N and S
layers are ${E}_{FN}= \hbar^2 k_{FN}^2/2m_N = \mu - U_N$ and
${E}_{FS}= \hbar^2 k_{FS}^2/2m_S =\mu-U_S$, respectively. We
define the corresponding effective chemical potentials as $\mu_N =
\mu_N(k_\parallel)={E}_{FN}-(\hbar^2 k_\parallel^2/2m_N)$ and
$\mu_S=\mu_S(k_\parallel)={E}_{FS}-(\hbar^2 k_\parallel^2/2m_S)$,
where ${\bf k}_\parallel$ is the conserved quasiparticle momentum
parallel to the layers. %\cite{art:Gyorffy96}.

\MyTextIt{The pair potential $\Delta({\bf r})$ should be treated
self-consis\-tent\-ly. For the sake of simplicity we used the
stepwise model with $\Delta({\bf r})$ equal to constant $\Delta_0$
in S and zero in N layers \cite{art:Kummel94,art:TanakaTsukada1}.
However, for S/N superlattices with thin S layers, the effective
$\Delta_0$ can be taken as the space-averaged self-consistently
determined pair potential, correspondingly smaller than the bulk
value \cite{art:BozovicRadovicNew}. For superlattices with thick S
films, $\Delta_0$ can be set to the bulk value.}

Solutions of BdG equation,
%%%%%%%%%%%%%%%%%%%%%%%%%%%%%%%%%%%%%%%%%%%%%%%%%%%%%%%%%%%%%%%%%%%%%%%%%%%%%
\begin{equation}
\Psi({\bf r})=
    \begin{pmatrix}
    u(z) \\ v(z)
    \end{pmatrix}
    e^{i{\bf k}_\parallel \cdot {\bf r}},
\end{equation}
%%%%%%%%%%%%%%%%%%%%%%%%%%%%%%%%%%%%%%%%%%%%%%%%%%%%%%%%%%%%%%%%%%%%%%%%%%%%%
in N and S layers can be written in the form
%%%%%%%%%%%%%%%%%%%%%%%%%%%%%%%%%%%%%%%%%%%%%%%%%%%%%%%%%%%%%%%%%%%%%%%%%%%%%
\begin{multline}
    \begin{pmatrix}
    u(z) \\ v(z)
    \end{pmatrix}_N
    =
    C_1 \sin(k^+_N z)
    \begin{pmatrix}
    1 \\ 0
    \end{pmatrix}
    +
    C_2 \cos(k^+_N z)
    \begin{pmatrix}
    1 \\ 0
    \end{pmatrix}
    \\
    +
    C_3 \sin(k^-_N z)
    \begin{pmatrix}
    0 \\ 1
    \end{pmatrix}
    +
    C_4 \cos(k^-_N z)
    \begin{pmatrix}
    0 \\ 1
    \end{pmatrix}
\end{multline}
and
%%%%%%%%%%%%%%%%%%%%%%%%%%%%%%%%%%%%%%%%%%%%%%%%%%%%%%%%%%%%%%%%%%%%%%%%%%%%%
\begin{multline}
    \begin{pmatrix}
    u(z) \\ v(z)
    \end{pmatrix}_S
    =
    C_5 \sin(k^+_S z)
    \begin{pmatrix}
    \bar u \\ \bar v
    \end{pmatrix}
    +
    C_6 \cos(k^+_S z)
    \begin{pmatrix}
    \bar u \\ \bar v
    \end{pmatrix}
    \\
    +
    C_7 \sin(k^-_S z)
    \begin{pmatrix}
    \bar v \\ \bar u
    \end{pmatrix}
    +
    C_8 \cos(k^-_S z)
    \begin{pmatrix}
    \bar v \\ \bar u
    \end{pmatrix}.
\end{multline}
%%%%%%%%%%%%%%%%%%%%%%%%%%%%%%%%%%%%%%%%%%%%%%%%%%%%%%%%%%%%%%%%%%%%%%%%%%%%%
Here,
$\Omega = \sqrt{\smash[b]{E^2 - \Delta_0^2}}$,
$k^\pm_N=\sqrt{\smash[b]{2m_N(\mu_N \pm E)/ \hbar^2}}$,
$k^\pm_S=\sqrt{\smash[b]{2m_S(\mu_S \pm \Omega)/ \hbar^2}}$ and the
BCS coherence amplitudes are
$\bar u = \sqrt{\smash[b]{(1+\Omega/E)/2}}$ and
$\bar v = \sqrt{\smash[b]{(1-\Omega/E)/2}}$.

Complex coefficients $C_1$ through $C_8$ are determined from the
boundary conditions at interfaces $z=0$ and $z=-d_N$ inside the primitive cell
%%%%%%%%%%%%%%%%%%%%%%%%%%%%%%%%%%%%%%%%%%%%%%%%%%%%%%%%%%%%%%%%%%%%%%%%%%%%%
\begin{equation}\label{eq:bCond1}
    \begin{pmatrix}
    u_N(0) \\ v_N(0)
    \end{pmatrix}
    =
    \begin{pmatrix}
    u_S(0) \\ v_S(0)
    \end{pmatrix},
\end{equation}
%%%%%%%%%%%%%%%%%%%%%%%%%%%%%%%%%%%%%%%%%%%%%%%%%%%%%%%%%%%%%%%%%%%%%%%%%%%%%
\begin{equation}\label{eq:bCond2}
    \frac{1}{m_N}
    \begin{pmatrix}
    u'_N(0) \\ v'_N(0)
    \end{pmatrix}
    +
    \frac{k_{FS}}{m_S} Z
    \begin{pmatrix}
    u(0) \\ v(0)
    \end{pmatrix}
    =
    \frac{1}{m_S}
    \begin{pmatrix}
    u'_S(0) \\ v'_S(0)
    \end{pmatrix},
\end{equation}
%%%%%%%%%%%%%%%%%%%%%%%%%%%%%%%%%%%%%%%%%%%%%%%%%%%%%%%%%%%%%%%%%%%%%%%%%%%%%
\begin{equation}\label{eq:bCond3}
    e^{iKa}
    \begin{pmatrix}
    u_N(-d_N) \\ v_N(-d_N)
    \end{pmatrix}
    =
    \begin{pmatrix}
    u_S(d_S) \\ v_S(d_S)
    \end{pmatrix},
\end{equation}
%%%%%%%%%%%%%%%%%%%%%%%%%%%%%%%%%%%%%%%%%%%%%%%%%%%%%%%%%%%%%%%%%%%%%%%%%%%%%
\begin{multline}\label{eq:bCond4}
    \frac{e^{iKa}}{m_N}
    \begin{pmatrix}
    u'_N(-d_N) \\ v'_N(-d_N)
    \end{pmatrix}
    -
    \frac{k_{FS}}{m_S} Z e^{iKa}
    \begin{pmatrix}
    u_N(-d_N) \\ v_N(-d_N)
    \end{pmatrix}
    \\
    =
    \frac{1}{m_S}
    \begin{pmatrix}
    u'_S(d_S) \\ v'_S(d_S)
    \end{pmatrix}.
\end{multline}
%%%%%%%%%%%%%%%%%%%%%%%%%%%%%%%%%%%%%%%%%%%%%%%%%%%%%%%%%%%%%%%%%%%%%%%%%%%%%
Here, the Bloch condition $\Psi(x,y,z+a)=e^{iKa}\Psi(x,y,z)$ is
used and the crystal momentum $K$ is taken within the first
Brillouin zone, $K \in (-\pi/a, \pi/a)$.

Dispersion relation $E=E_{n,k_\parallel}(K)$ can be
written in the following implicit form \cite{art:Kummel94}
%%%%%%%%%%%%%%%%%%%%%%%%%%%%%%%%%%%%%%%%%%%%%%%%%%%%%%%%%%%%%%%%%%%%%%%%%%%%%
\begin{align}\label{eq:DispRel}
\cos(K a) &= - \tilde D_1 /4 \pm
\sqrt{(\tilde D_1/4)^2 - \tilde D_2/4 + 1/2} \notag
\\
&\equiv F^\pm(E,k_\parallel),
\end{align}
%%%%%%%%%%%%%%%%%%%%%%%%%%%%%%%%%%%%%%%%%%%%%%%%%%%%%%%%%%%%%%%%%%%%%%%%%%%%%
where $\tilde D_1$ and $\tilde D_2$ are defined in terms of dimensionless quantities
%%%%%%%%%%%%%%%%%%%%%%%%%%%%%%%%%%%%%%%%%%%%%%%%%%%%%%%%%%%%%%%%%%%%%%%%%%%%%
$E/\Delta_0$,
$k_\parallel / k_{FS}$,
$Z$,
$d_N/ \xi_0$,
$d_S / \xi_0$,
$m_N / m_S$,\\
${E}_{FN}/{E}_{FS}$, and
$\Delta_0 / {E}_{FS}$
%%%%%%%%%%%%%%%%%%%%%%%%%%%%%%%%%%%%%%%%%%%%%%%%%%%%%%%%%%%%%%%%%%%%%%%%%%%%%
(see the Appendix).

Global density of states (for both spin orientations) per unit
area of the cross section $L_xL_y$, averaged over a primitive cell,
is given by
%%%%%%%%%%%%%%%%%%%%%%%%%%%%%%%%%%%%%%%%%%%%%%%%%%%%%%%%%%%%%%%%%%%%%%%%%%%%%
\begin{align}\label{eq:g}
g(E)&=
    \frac{1}{L_xL_y} \sum_{\sigma,k_\parallel, K}
    \delta {\bm (} E-E(k_\parallel,K) {\bm )} \notag \\
    &=
    \frac{1}{\pi}\int dk_\parallel \; k_\parallel
    \sum_{i=+,-} \frac{a}{2\pi} \int dK^i \;
    \delta {\bm (} E-E(k_\parallel,K^i) {\bm )} \notag \\
    &=
    \frac{a}{2\pi^2} \int dk_\parallel \; k_\parallel
    \sum_{i=+,-}
    \left|
    \frac{\partial K^i}{\partial E(k_\parallel,K^i)}
    \right|_{E(k_\parallel, K^i)=E},
\end{align}
%%%%%%%%%%%%%%%%%%%%%%%%%%%%%%%%%%%%%%%%%%%%%%%%%%%%%%%%%%%%%%%%%%%%%%%%%%%%%
where $K^\pm(E)$ are the solutions of Eq. \eqref{eq:DispRel}, and
%%%%%%%%%%%%%%%%%%%%%%%%%%%%%%%%%%%%%%%%%%%%%%%%%%%%%%%%%%%%%%%%%%%%%%%%%%%%%
\begin{equation}
\left|
\frac{\partial K^i}{\partial E(k_\parallel,K^i)}
\right|
=
\frac{1}{a}
\left|
\frac{\partial\arccos[F^i(E,k_\parallel)]}{\partial E}
\right|.
\end{equation}
%%%%%%%%%%%%%%%%%%%%%%%%%%%%%%%%%%%%%%%%%%%%%%%%%%%%%%%%%%%%%%%%%%%%%%%%%%%%%
In accordance with Eq. \eqref{eq:DispRel}, the integration over
$k_\parallel$ [or $\mu_S=\mu_S(k_\parallel)$] in Eq. \eqref{eq:g}
is limited to the intervals given by
%%%%%%%%%%%%%%%%%%%%%%%%%%%%%%%%%%%%%%%%%%%%%%%%%%%%%%%%%%%%%%%%%%%%%%%%%%%%%
\begin{equation}\label{eq:Cond1}
(\tilde D_1/4)^2 - \tilde D_2/4 + 1/2 \ge 0
\end{equation}
%%%%%%%%%%%%%%%%%%%%%%%%%%%%%%%%%%%%%%%%%%%%%%%%%%%%%%%%%%%%%%%%%%%%%%%%%%%%%
and
%%%%%%%%%%%%%%%%%%%%%%%%%%%%%%%%%%%%%%%%%%%%%%%%%%%%%%%%%%%%%%%%%%%%%%%%%%%%%
\begin{equation}\label{eq:Cond2}
|F^\pm(E,k_\parallel)| \le 1.
\end{equation}
%%%%%%%%%%%%%%%%%%%%%%%%%%%%%%%%%%%%%%%%%%%%%%%%%%%%%%%%%%%%%%%%%%%%%%%%%%%%%
In the following, $g(E)$ is normalized to the normal-state value
$\bar g = (m_S d_S k_{FS} + m_N d_N k_{FN})/ \pi^2\hbar^2$.

%%%%%%%%%%%%%%%%%%%%%%%%%%%%%%%%%%%%%%%%%%%%%%%%%%%%%%%%%%%%%%%%%%%%%%%%%%%%%
%%%%%%%%%%%%%%%%%%%%%%%%%%%%%%%%%%%%%%%%%%%%%%%%%%%%%%%%%%%%%%%%%%%%%%%%%%%%%
\section{\label{sec:Elect}Energy bands and density of states}
%\protect\\

%%%%%%%%%%%%%%%%%%%%%%%%%%%%%%%%%%%%%%%%%%%%%%%%%%%%%%%%%%%%%%%%%%%%%%%%%%%%%
%%%%%%%%%%%%%%%%%%%%%%%%%%%%%%%%%%%%%%%%%%%%%%%%%%%%%%%%%%%%%%%%%%%%%%%%%%%%%
\begin{figure}[t]
%FIG 1
\vspace*{3mm}
\includegraphics[width=7.7cm, height=6.3cm]{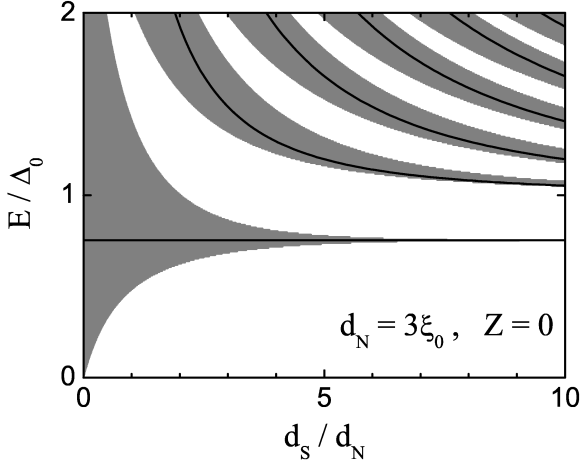} %EvsDZ0.EPS
\caption{\label{fig:EvsDZ0} Energy bands as a function of the S layer thickness
            $d_S$, for S/N superlattices
            with thick N layers, $d_N=3\xi_0$, transparent interfaces, $Z=0$,
            and ${\bf k}_\parallel=0$. Andreev
            bound states ($E<\Delta_0$) and geometrical resonances ($E>\Delta_0$)
            for the corresponding SNS and NSN
            trilayers are shown for comparison (solid curves).}
\end{figure}
%%%%%%%%%%%%%%%%%%%%%%%%%%%%%%%%%%%%%%%%%%%%%%%%%%%%%%%%%%%%%%%%%%%%%%%%%%%%%
%%%%%%%%%%%%%%%%%%%%%%%%%%%%%%%%%%%%%%%%%%%%%%%%%%%%%%%%%%%%%%%%%%%%%%%%%%%%%
\begin{figure}[t]
%FIG 2
\vspace*{3mm}
\includegraphics[width=8cm, height=6.3cm]{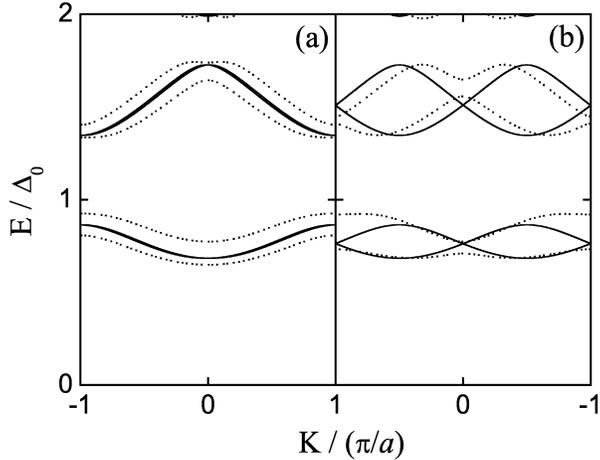} %EvsK12Z1.EPS
\caption{\label{fig:EvsK12} Characteristic dispersion of energy
            bands illustrated for $d_N=3\xi_0$, $Z=0$,
            ${\bf k}_\parallel=0$, and (a) $d_S=3.0005 d_N$, and
            (b) $d_S=3.0013d_N$. Bands displayed in (a) are double
            degenerate for all $K$. Dotted curves represent dispersion
            for finite transparency $Z=0.5$.}
\end{figure}
%%%%%%%%%%%%%%%%%%%%%%%%%%%%%%%%%%%%%%%%%%%%%%%%%%%%%%%%%%%%%%%%%%%%%%%%%%%%%
%%%%%%%%%%%%%%%%%%%%%%%%%%%%%%%%%%%%%%%%%%%%%%%%%%%%%%%%%%%%%%%%%%%%%%%%%%%%%

The dispersion relation, Eq. \eqref{eq:DispRel}, is solved
numerically and the global density of states is calculated from Eq.
\eqref{eq:g} for various superlattices and for zero phase difference
$\phi=0$. In the following, we focus on the influence of finite
interface transparency on quasiparticle band structure and density
of states. For simplicity, this is illustrated for equal effective
masses and Fermi wave-vectors, $m_N/m_S=1$ and $k_{FN}/k_{FS}=1$.
Superconductors are characterized by \MyTextIt{the bulk value of the
pair potential} $\Delta_0/E_{FS}=10^{-3}$, which corresponds to the
zero-tem\-p\-er\-ature BCS coherence length $\xi_0=\hbar^2
k_{FS}/(\pi m_S \Delta_0)$ $\sim 10^3$\AA.

Energy bands for S/N superlattices with thick N layers,
transparent S-N interfaces and quasiparticles propagating
perpendicular to the layers (${\bf k}_\parallel=0$) are shown in
Fig. \ref{fig:EvsDZ0}.
Quasicontinuum of energy states corresponding to the crystal momentum
within the first Brillouin zone, $K\in(-\pi/a,\pi/a)$, is
indicated by shading the band width calculated from Eqs. \eqref{eq:Cond1} and
\eqref{eq:Cond2}.

%%%%%%%%%%%%%%%%%%%%%%%%%%%%%%%%%%%%%%%%%%%%%%%%%%%%%%%%%%%%%%%%%%%%%%%%%%%%%
%%%%%%%%%%%%%%%%%%%%%%%%%%%%%%%%%%%%%%%%%%%%%%%%%%%%%%%%%%%%%%%%%%%%%%%%%%%%%
\begin{figure}[t]
%FIG 3
\vspace*{3mm}
\hspace*{7mm}
\includegraphics[width=6.84cm, height=8.4cm]{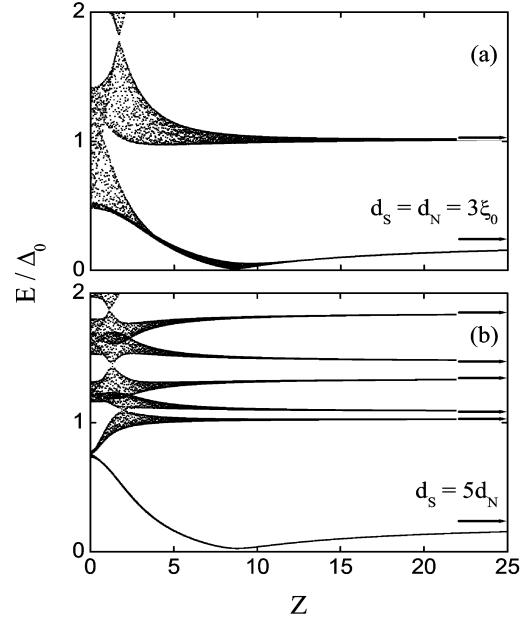} %EvsZd12.EPS
\caption{\label{fig:EvsZd12} Energy bands as a function of $Z$, for
            ${\bf k}_\parallel=0$ and for two particular S/N superlattices:
            (a) $d_S=d_N=3\xi_0$ and (b) $d_S=15\xi_0$,
            $d_N=3\xi_0$. Arrows indicate the bound states in the tunnel limit.}
\end{figure}
%%%%%%%%%%%%%%%%%%%%%%%%%%%%%%%%%%%%%%%%%%%%%%%%%%%%%%%%%%%%%%%%%%%%%%%%%%%%%
%%%%%%%%%%%%%%%%%%%%%%%%%%%%%%%%%%%%%%%%%%%%%%%%%%%%%%%%%%%%%%%%%%%%%%%%%%%%%
For the corresponding SNS trilayer, Andreev bound states, $E<\Delta_0$,
in the normal interlayer of thickness $d_N$, for zero phase difference
across the junction, transparent interfaces,  and ${\bf k}_\parallel=0$
are given by \cite{art:Kulik}
%\cite{art:Baselmans} \cite{art:GunsenhKummel94,art:Kulik}
%%%%%%%%%%%%%%%%%%%%%%%%%%%%%%%%%%%%%%%%%%%%%%%%%%%%%%%%%%%%%%%%%%%%%%%%%%%%%
\begin{equation}\label{eq:SNSAndreev}
\frac{E_n}{\Delta_0} = \pi^2
\Big[
n+ \frac{1}{\pi} \arccos\Big(\frac{E_n}{\Delta_0}\Big)
\Big]
\frac{1}{d_N/\xi_0},
\end{equation}
%%%%%%%%%%%%%%%%%%%%%%%%%%%%%%%%%%%%%%%%%%%%%%%%%%%%%%%%%%%%%%%%%%%%%%%%%%%%%
where $n=0,1,\ldots$. In this case the Andreev bound states are double
degenerate.
Geometrical resonances, $E>\Delta_0$, for the
corresponding NSN junction with S interlayer of thickness $d_S$,
and $Z=0$, ${\bf k}_\parallel=0$
are given by
%%%%%%%%%%%%%%%%%%%%%%%%%%%%%%%%%%%%%%%%%%%%%%%%%%%%%%%%%%%%%%%%%%%%%%%%%%%%%
\begin{equation}\label{eq:NSNGeom}
\frac{E_n}{\Delta_0}= \sqrt{1+n^2\frac{\pi^4}{(d_S/\xi_0)^2}},
\end{equation}
%%%%%%%%%%%%%%%%%%%%%%%%%%%%%%%%%%%%%%%%%%%%%%%%%%%%%%%%%%%%%%%%%%%%%%%%%%%%%
which follows from the condition $d_S(k^+_S - k^-_S)=2n\pi$, where
$n=\pm 1,\pm 2,\ldots$. At these energies the Andreev reflection
vanishes and the electron is transmitted without creation or
annihilation of Cooper pairs
\cite{art:BozovicRadovic,art:BozovicRadovicNew,art:RadovicPhysC}.
Both Andreev bound states and geometrical resonances of the
corresponding SNS and NSN trilayers are shown in Fig.
\ref{fig:EvsDZ0} for comparison.

In S/N superlattices with thick S layers, the energy band structure
above the gap, $E > \Delta_0$, is also affected by the Andreev
process \cite{art:Gyorffy96,art:GunsenhKummel94}. With increasing
$d_N$, the band structure dependance on $d_S/d_N$ remains
qualitatively the same as in Fig. \ref{fig:EvsDZ0}, with compression
and lowering of energy bands that enter the superconducting gap
\cite{art:Vanevic}. \MyTextIt{Andreev reflection is the fundamental
mechanism that determines the quasiparticle band structure in S/N
superlattices. However, qualitatively the same results as shown in
Fig. \ref{fig:EvsDZ0} are obtained for semiconductor / normal-metal
superlattices \cite{art:SurfSciRep,art:PikusBook}.} Characteristic
dispersion of energy ba\-nds, shown in Fig. \ref{fig:EvsDZ0}, is
illustrated in Fig. \ref{fig:EvsK12} for two close thicknesses of
the S layer. For some layer thicknesses the energy bands are double
degenerate for all $K$, Fig. \ref{fig:EvsK12} (a), in contrast with
the usual degeneracy at high-symmetry points only (at the center and
the ends of the first Brillouin zone), Fig. \ref{fig:EvsK12} (b).
These two types of dispersion alternate rapidly with the change of
layer thicknesses on the atomic scale $k_{F}^{-1}$, while the band
width changes on the macroscopic scale.

\begin{figure}[t]
% FIG 4
\hfuzz15pt
\vspace*{3mm}
\hspace*{2mm}
\includegraphics[width=8.5cm, height=6.4cm]{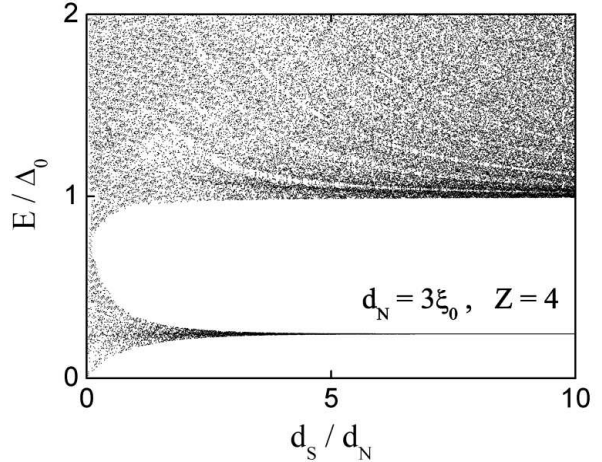} %EvsDZ1-1
\caption{\label{fig:EvsDZ1} Energy 'bands' for $d_N=3\xi_0$,
            ${\bf k}_\parallel=0$ and $Z=4$. Shading is produced by the rapid
            oscillatory dependence of flat energy bands on the S layer
            thickness.}
\end{figure}

\begin{figure}[t]
% FIG 5
\hfuzz15pt
\vspace*{5mm}
\hspace*{2mm}
\includegraphics[width=7.7cm, height=6.2cm]{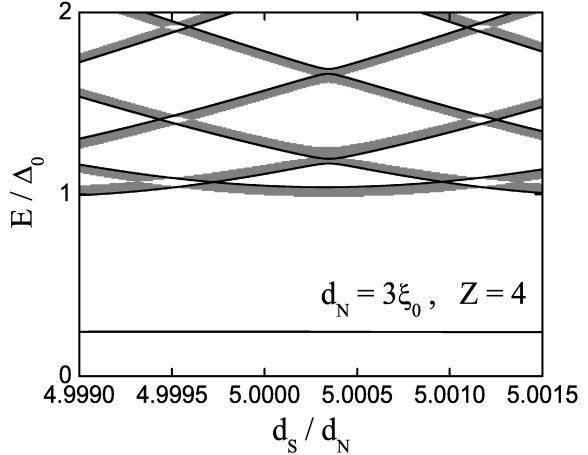} %EvsDZ1close
\caption{\label{fig:EvsDZ1close} Rapid oscillatory dependance of
            energy bands on the S layer thickness for nontransparent
            superlattices. Energy dispersion $E(K)$
            is shown by shading for all $K$, solid curves represent $E(0)$.}
\end{figure}

Finite interface transparency, as well as mismatch of effective
masses and \MyTextIt{Fermi wave-vectors} \cite{art:Vanevic}, lift the
degeneracy in $E(K)$, Fig. \ref{fig:EvsK12},  and change the band
structure, Fig. \ref{fig:EvsZd12}. For large $Z$, energy bands split
into pairs of flat bands independent of $K$, and there is a
significant change of the band energy below the superconducting gap.
Approaching the tunnel limit for $Z\gg 1$, pairs of adjacent flat
energy bands transform into bound states of isolated films defined
by $d_S k^\pm_S = n_1\pi$ and $d_N k^\pm_N = n_2\pi$. However, this
does not imply that the energy band splitting and decrease of the
band widths due to the flattening will be visible in the $E$ vs.
$d_{S(N)}$ plot (cf. Figs. \ref{fig:EvsDZ0} and \ref{fig:EvsDZ1}).
Energy levels for $Z\gg 1$ oscillate rapidly with layer thicknesses
on the atomic scale $k_F^{-1}$, Fig. \ref{fig:EvsDZ1close}, so that
$E$ vs. $d_{S(N)}$ curves fill the energy space quasicontinuously on
the macroscopic scale, Fig. \ref{fig:EvsDZ1}. This implies erasing
of the band structure and localization of quasiparticle states in
real superlattices with finite interface transparency and slightly
unequal layers.
%%%%%%%%%%%%%%%%%%%%%%%%%%%%%%%%%%%%%%%%%%%%%%%%%%%%%%%%%%%%%%%%%%%%%%%%%%%%%
%%%%%%%%%%%%%%%%%%%%%%%%%%%%%%%%%%%%%%%%%%%%%%%%%%%%%%%%%%%%%%%%%%%%%%%%%%%%%
%%%%%%%%%%%%%%%%%%%%%%%%%%%%%%%%%%%%%%%%%%%%%%%%%%%%%%%%%%%%%%%%%%%%%%%%%%%%%
%%%%%%%%%%%%%%%%%%%%%%%%%%%%%%%%%%%%%%%%%%%%%%%%%%%%%%%%%%%%%%%%%%%%%%%%%%%%%
%%%%%%%%%%%%%%%%%%%%%%%%%%%%%%%%%%%%%%%%%%%%%%%%%%%%%%%%%%%%%%%%%%%%%%%%%%%%%
%%%%%%%%%%%%%%%%%%%%%%%%%%%%%%%%%%%%%%%%%%%%%%%%%%%%%%%%%%%%%%%%%%%%%%%%%%%%%

\begin{figure}[t]
%FIG 6
\vspace*{3mm}
\includegraphics[width=7.7cm, height=6.3cm]{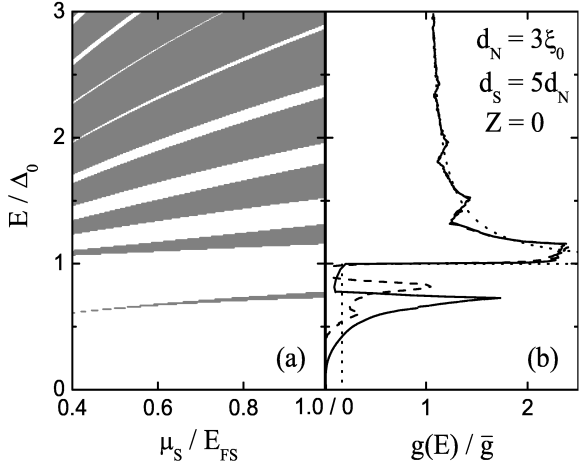} %d1Z0debeli12-sviAB.EPS
\caption{\label{fig:d1Z0debeli12-svi} (a) \MyTextIt{Normalized energy bands
            $E/\Delta_0$} as a function of $\mu_S=\mu_S(k_\parallel)$, for S/N
            superlattice
            with $d_S=15\xi_0$, $d_N=3\xi_0$ and $Z=0$,
            and  (b) \MyTextIt{the corresponding global density of states $g(E)$
            normalized to the normal-state value $\overline{g}$.}
            Global density of states for
            $Z=0.5$ (dashed curve), and in the tunnel
            limit (dotted curve) are given for comparison.}
\end{figure}

\begin{figure}[t]
%FIG 7
\vspace*{3mm}
\includegraphics[width=7.7cm, height=6.3cm]{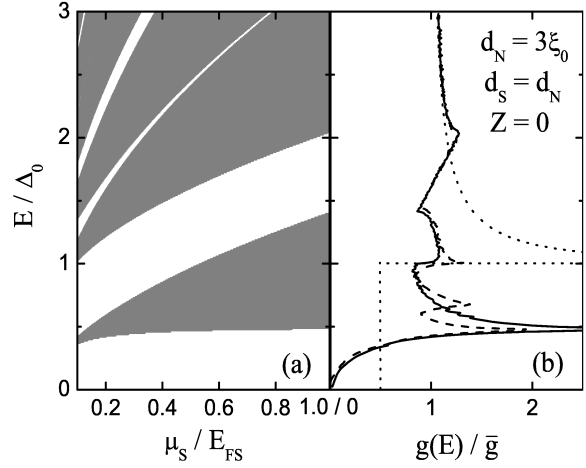} %d2Z0debeli12-sviAB.EPS
\caption{\label{fig:d2Z0debeli12-svi} (a) \MyTextIt{Normalized energy bands
            $E/\Delta_0$ } as a function of
            $\mu_S=\mu_S(k_\parallel)$, for S/N superlattice
            with $d_S=d_N=3\xi_0$ and $Z=0$, and
            (b) \MyTextIt{the corresponding global density of states $g(E)$
            normalized to the normal-state value $\overline{g}$.}
            Global density of states for
            $Z=0.5$ (dashed curve), and in the tunnel
            limit (dotted curve) are given for comparison.}
\end{figure}

Previous analysis has been made for quasiparticles that propagate
perpendicular to the layers. Dependence of energy bands on $k_\parallel$,
i.e. on the effective chemical potential $\mu_S(k_\parallel)$,
is illustrated for $Z=0$ in Figs. \ref{fig:d1Z0debeli12-svi} (a)
and \ref{fig:d2Z0debeli12-svi} (a).
Band widths decrease with the increase of $k_\parallel$, and bands split
into pairs of bound states (flat bands) for very large parallel
momentum \cite{art:Kummel94}, similar to the tunnel limit.
The effect of erasing the band structure with finite interface transparency
is enhanced with the increase of $k_\parallel$. Corresponding changes of the global
density of states are shown in Figs. \ref{fig:d1Z0debeli12-svi} (b)
and \ref{fig:d2Z0debeli12-svi} (b).  Integration of Eq. \eqref{eq:g} is
performed over the shaded regions in Figs. \ref{fig:d1Z0debeli12-svi} (a)
and \ref{fig:d2Z0debeli12-svi} (a), where Eqs. \eqref{eq:Cond1} and
\eqref{eq:Cond2} are satisfied.

\begin{figure}[b]
% FIG 8
\vspace*{3mm}
\hspace*{7mm}
\includegraphics[width=6.84cm, height=8.4cm]{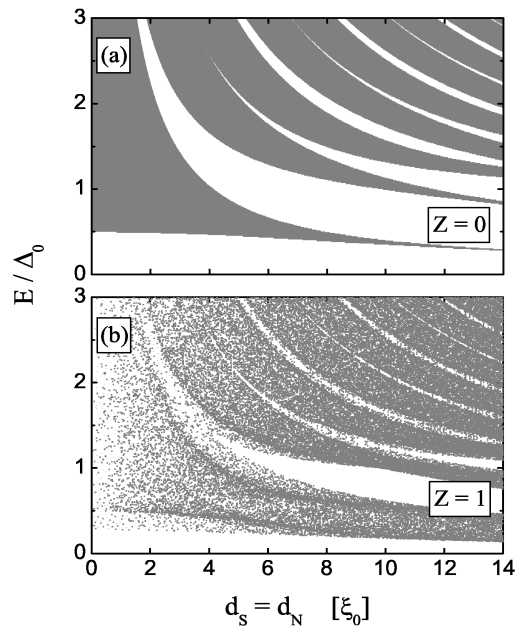} %EvsDjedZ12.EPS
\caption{\label{fig:EvsD} (a) Energy bands for S/N superlattices with
            $d_S=d_N$, ${\bf k}_\parallel=0$ and $Z=0$.
            (b) Erasing of the band structure with decrease of interface
            transparency is shown for $Z=1$.}
\end{figure}

\begin{figure}[t]
% FIG 9
\vspace*{3mm}
\hspace*{7mm}
\includegraphics[width=5.7cm, height=13.4cm]{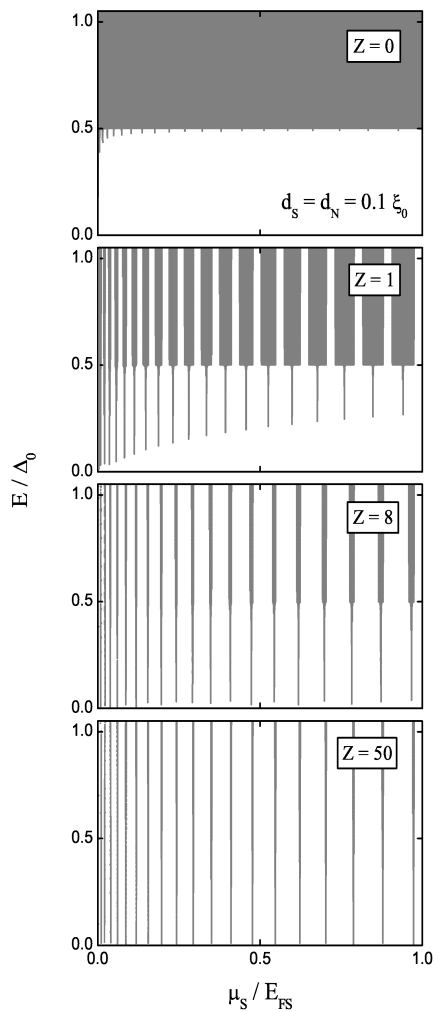} %EvsMiSveD1.EPS
\caption{\label{fig:EvsMiSveD1} Energy bands as a function of
            $\mu_S=\mu_S(k_\parallel)$, for S/N superlattice with
            thin \MyTextIt{layers}, $d_S=d_N=0.1\xi_0$, and for various interface transparencies.}
\end{figure}

\begin{figure}[t]
% FIG 10
\vspace*{2mm}
\hspace*{7mm}
\includegraphics[width=5.7cm, height=13.4cm]{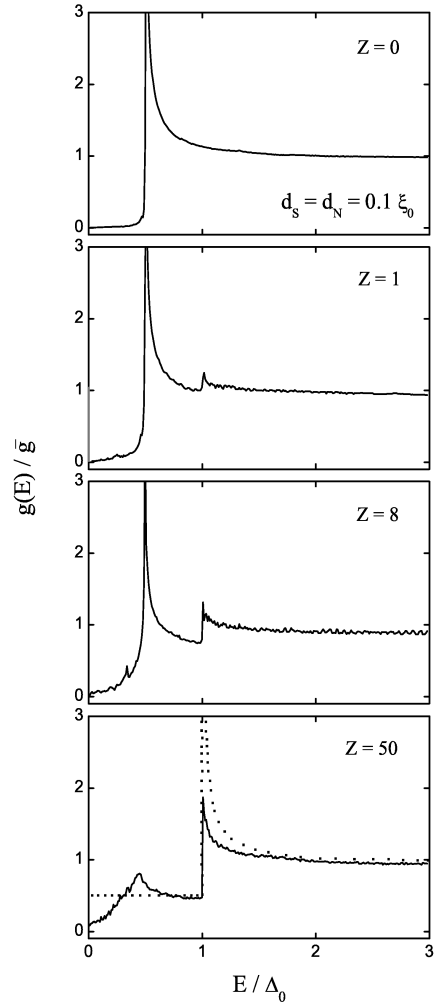} %gustZsvi-2.EPS
\vspace*{1mm}
\caption{\label{fig:gustZsvi} Global density of states for S/N
            superlattice with thin \MyTextIt{layers}, $d_S=d_N=0.1\xi_0$, and for various
            interface transparencies. Tunnel limit is indicated in the bottom
            panel (dotted curve). Note that \MyTextIt{the effective} $\Delta_0$ varies with $Z$, being the
            smallest for $Z=0$ (top panel) and reaching the bulk
            value in the tunnel limit (bottom panel).}
\end{figure}

Now we shall focus on energy bands and the density of states in
thin-layer S/N superlattices, where coherence effects are
pronounced and ballistic transport is more likely to take place
\cite{art:RadovicPhysC}. Dependence of energy bands on the
superlattice period is illustrated in Fig. \ref{fig:EvsD} for
$d_S=d_N$, ${\bf k}_\parallel=0$, and for both $Z=0$ and $Z=1$. It
can be seen that the band structure in transparent thin-layer
superlattices differs significantly from the thick-layer case
considered in Ref. \cite{art:Kummel94}. For thin layers,
dispersion of energy bands is significant, with only a small part
of the lower band laying below $\Delta_0$. Energy bands as a
function of $k_\parallel$, and for various interface
transparencies are shown in Fig. \ref{fig:EvsMiSveD1}. For $Z=0$
and $d_S=d_N$, the most striking feature is the onset of the
lowest energy band at the midgap, practically for any
$k_\parallel$. This is not the case for thick-layer superlattices,
where band energy decreases more rapidly down to zero with the
increase of $k_\parallel$, resulting in the left-side "tail" of
the subgap peak in the density of states, Figs.
\ref{fig:d1Z0debeli12-svi} and \ref{fig:d2Z0debeli12-svi}. For
thin-layer superlattices, finite interface transparency introduces
the resonance effect: energy bands penetrate periodically below
the midgap with the increase of $k_\parallel$. This is more
pronounced as $Z$ gets larger, Fig. \ref{fig:EvsMiSveD1}. The
corresponding global densities of states for various interface
transparencies are shown in Fig. \ref{fig:gustZsvi}.
%%%%%%%%%%%%%%%%%%%%%%%%%%%%%%%%%%%%%%%%%%%%%%%%%%%%%%%%%%%%%%%%%%%%%%%%%%%%%
%%%%%%%%%%%%%%%%%%%%%%%%%%%%%%%%%%%%%%%%%%%%%%%%%%%%%%%%%%%%%%%%%%%%%%%%%%%%%
%%%%%%%%%%%%%%%%%%%%%%%%%%%%%%%%%%%%%%%%%%%%%%%%%%%%%%%%%%%%%%%%%%%%%%%%%%%%%
%%%%%%%%%%%%%%%%%%%%%%%%%%%%%%%%%%%%%%%%%%%%%%%%%%%%%%%%%%%%%%%%%%%%%%%%%%%%%

For transparent interfaces, the density of states is BCS-like with
the energy gap $E_g$ smaller than the pair potential $\Delta_0$.
The value of $E_g$ for transparent interfaces, equal effective
masses, and equal Fermi energies can be obtained from the well
known dispersion relation
\cite{art:TanakaTsukada,art:vanGelder,art:Kummel71}
%%%%%%%%%%%%%%%%%%%%%%%%%%%%%%%%%%%%%%%%%%%%%%%%%%%%%%%%%%%%%%%%%%%%%%%%%%%%%
\begin{multline}\label{eq:vanGelder}
\cos [ (K^\pm \pm  k_{zF})a ]
=
\cos(q\delta \, d_S)\cos(q\, d_N)
\\
-
\delta^{-1} \sin(q\delta \, d_S)\sin(q\, d_N),
\end{multline}
%%%%%%%%%%%%%%%%%%%%%%%%%%%%%%%%%%%%%%%%%%%%%%%%%%%%%%%%%%%%%%%%%%%%%%%%%%%%%
which is a special case of Eq. \eqref{eq:DispRel}. Here,
$k_{zF}=\sqrt{\smash[b]{k_F^2 - k_\parallel^2}}$,
$\delta=\Omega/E$, and $q=mE/\hbar^2 k_{zF}$. For $d_S$, $d_N \to
0$, from Eqs. \eqref{eq:g} and \eqref{eq:vanGelder} exactly
follows \cite{art:BaranovBuzdin}
%%%%%%%%%%%%%%%%%%%%%%%%%%%%%%%%%%%%%%%%%%%%%%%%%%%%%%%%%%%%%%%%%%%%%%%%%%%%%
\begin{equation}
E_g=\frac{\Delta_0}{1+d_N/d_S}.
\end{equation}
%%%%%%%%%%%%%%%%%%%%%%%%%%%%%%%%%%%%%%%%%%%%%%%%%%%%%%%%%%%%%%%%%%%%%%%%%%%%%
Practically, this simple relation remains valid for the layer
thicknesses up to one or two coherence lengths $\xi_0$, due to
the weak variation of the bottom of the lowest energy band with the
layer thickness, Fig. \ref{fig:EvsD} (a).
%%%%%%%%%%%%%%%%%%%%%%%%%%%%%%%%%%%%%%%%%%%%%%%%%%%%%%%%%%%%%%%%%%%%%%%%%%%%%
%%%%%%%%%%%%%%%%%%%%%%%%%%%%%%%%%%%%%%%%%%%%%%%%%%%%%%%%%%%%%%%%%%%%%%%%%%%%%
%%%%%%%%%%%%%%%%%%%%%%%%%%%%%%%%%%%%%%%%%%%%%%%%%%%%%%%%%%%%%%%%%%%%%%%%%%%%%
%%%%%%%%%%%%%%%%%%%%%%%%%%%%%%%%%%%%%%%%%%%%%%%%%%%%%%%%%%%%%%%%%%%%%%%%%%%%%
%%%%%%%%%%%%%%%%%%%%%%%%%%%%%%%%%%%%%%%%%%%%%%%%%%%%%%%%%%%%%%%%%%%%%%%%%%%%%
%%%%%%%%%%%%%%%%%%%%%%%%%%%%%%%%%%%%%%%%%%%%%%%%%%%%%%%%%%%%%%%%%%%%%%%%%%%%%

With decreasing interfacial transparency, the subgap peak in
$g(E)$ at $E_g$ decays, and the usual BCS coherence peak at
$\Delta_0$ reenters as the superconducting layers become more
isolated. In the tunnel limit, the BCS peak at $\Delta_0$ is
completely restored, Fig. \ref{fig:gustZsvi} (dotted curve in the
bottom panel). For thicker layers $d_S \sim d_N\sim\xi_0$, the
coherence effects are less pronounced and the tunnel limit
behavior is practically reached for smaller $Z\sim 1$. Previously,
this double peak structure in the density of states of S/N
superlattices is obtained within the tight binding approximation
for atomic-scale layered systems, and apparently observed in
high-$T_c$ intrinsically layered superconductors
\cite{art:Bulaevski90,art:Buzdin92,art:Buschmann92}.
%%%%%%%%%%%%%%%%%%%%%%%%%%%%%%%%%%%%%%%%%%%%%%%%%%%%%%%%%%%%%%%%%%%%%%%%%%%%%
%%%%%%%%%%%%%%%%%%%%%%%%%%%%%%%%%%%%%%%%%%%%%%%%%%%%%%%%%%%%%%%%%%%%%%%%%%%%%
%%%%%%%%%%%%%%%%%%%%%%%%%%%%%%%%%%%%%%%%%%%%%%%%%%%%%%%%%%%%%%%%%%%%%%%%%%%%%
%%%%%%%%%%%%%%%%%%%%%%%%%%%%%%%%%%%%%%%%%%%%%%%%%%%%%%%%%%%%%%%%%%%%%%%%%%%%%
%%%%%%%%%%%%%%%%%%%%%%%%%%%%%%%%%%%%%%%%%%%%%%%%%%%%%%%%%%%%%%%%%%%%%%%%%%%%%
%%%%%%%%%%%%%%%%%%%%%%%%%%%%%%%%%%%%%%%%%%%%%%%%%%%%%%%%%%%%%%%%%%%%%%%%%%%%%

%%%%%%%%%%%%%%%%%%%%%%%%%%%%%%%%%%%%%%%%%%%%%%%%%%%%%%%%%%%%%%%%%%%%%%%%%%%%%
%%%%%%%%%%%%%%%%%%%%%%%%%%%%%%%%%%%%%%%%%%%%%%%%%%%%%%%%%%%%%%%%%%%%%%%%%%%%%
\section{\label{sec:Concl}Conclusion}

We have derived the dispersion relation for superconductor /
normal-metal (semiconductor) superlattices in the clean limit,
generalizing the previous expression of Plehn {et al.}
\cite{art:Kummel94} to include an arbitrary interface transparency
and mismatch of effective band masses. The obtained general
dispersion relation is used for numerical analysis of the influence
of interface transparency on energy band structure and density of
states in metallic S/N superlattices. Although we used stepwise
approximation for the pair potential, our results will not be
altered significantly by the fully self-consistent numerical
calculations \cite{art:Kummel94}, if \MyTextIt{an effective} pair
potential (smaller than the bulk value) is taken for thin S layers,
and simply the bulk value in the cases of thick S layers, low
transparency, and mismatch of \MyTextIt{Fermi wave-vectors} or band
masses. Our results confirm previously obtained features in the
metallic S/N superlattices
\MyTextIt{\cite{art:TanakaTsukada,art:Stojkovic,art:Kummel94}},
including the limiting cases of double barrier SNS or NSN trilayers
\MyTextIt{\cite{art:SchusslerKummel93,art:GunsenhKummel94,art:Brinkman,%
art:RadovicGrcka,art:BozovicRadovic,art:BozovicRadovicNew,art:RadovicPhysC}},
and are in a good qualitative agreement with the results obtained
within the tight binding approximation
\cite{art:Bulaevski90,art:Buzdin92}.

Consequences of the quantum interference effect are strong and
rapid (on the atomic scale) geometrical oscillations with layers
thickness of the energy dispersion in transparent superlattices,
and of the almost flat energy bands (transmission resonances) in
the case of finite transparency. Oscillations in the latter case
are practically within the band width of the corresponding fully
transparent superlattice. Practically, this could imply the
localization of quasiparticle states in superlattices with low
interface transparency.

Characteristic changes of quasiparticle band structure with
decreasing interface transparency are suitably reflected in the
global (averaged on the lattice period) density of states, which can
be directly measured by STM techniques
\cite{art:Buschmann92,art:Moussy02}. Oscillations of the density of
states are simply related to the band structure for transparent
superlattices with thick layers \cite{art:TanakaTsukada}. However,
superlattices with the period smaller than the coherence length,
when S and N layers lose their individual properties due to strong
\MyTextIt{phase coupling by Andreev scattering}, do not differ
significantly from a bulk BCS superconductor, except for a subgap
peak in the global density of states instead of the superconductor
coherence peak at $\Delta_0$ \cite{art:Bulaevski90}. For transparent
interfaces, position of the subgap peak is simply related to the
lattice parameters. For finite interface transparency, we find the
characteristic double peak structure in the global density of states
\cite{art:Buzdin92}. With decreasing transparency the subgap peak
decreases, only slightly changing the position, while the
coherence peak at the effective $\Delta_0$ grows, and the density
of states develops towards the gapless BCS result for the bulk
superconductor in the tunnel limit. We point out that this
double-peak structure of the global density of states in
small-period clean-metal S/N superlattices can be used as a reliable
experimental probe for interface transparency.

%%%%%%%%%%%%%%%%%%%%%%%%%%%%%%%%%%%%%%%%%%%%%%%%%%%%%%%%%%%%%%%%%%%%%%%%%%%%%
%%%%%%%%%%%%%%%%%%%%%%%%%%%%%%%%%%%%%%%%%%%%%%%%%%%%%%%%%%%%%%%%%%%%%%%%%%%%%
%\section*{ACKNOWLEDGMENTS}

\begin{acknowledgement}
We are grateful to Ivana Petkovi\' c, Milo\v s Bo\v z\-o\-vi\' c,
and Boris Grbi\' c for useful discussions. The work has been
supported by the Serbian Ministry of Science, Project No. 1899.
\end{acknowledgement}

%%%%%%%%%%%%%%%%%%%%%%%%%%%%%%%%%%%%%%%%%%%%%%%%%%%%%%%%%%%%%%%%%%%%%%%%%%%%%
%%%%%%%%%%%%%%%%%%%%%%%%%%%%%%%%%%%%%%%%%%%%%%%%%%%%%%%%%%%%%%%%%%%%%%%%%%%%%

%%%%%%%%%%%%%%%%%%%%%%%%%%%%%%%%%%%%%%%%%%%%%%%%%%%%%%%%%%%%%%%%%%%%%%%%%%%%%
%%%%%%%%%%%%%%%%%%%%%%%%%%%%%%%%%%%%%%%%%%%%%%%%%%%%%%%%%%%%%%%%%%%%%%%%%%%%%

%\begin{widetext}

%%%%%%%%%%%%%%%%%%%%%%%%%%%%%%%%%%%%%%%%%%%%%%%%%%%%%%%%%%%%%%%%%%%%%%%%%%%%%
%%%%%%%%%%%%%%%%%%%%%%%%%%%%%%%%%%%%%%%%%%%%%%%%%%%%%%%%%%%%%%%%%%%%%%%%%%%%%
%%%%%%%%%%%%%%%%%%%%%%%%%%%%%%%%%%%%%%%%%%%%%%%%%%%%%%%%%%%%%%%%%%%%%%%%%%%%%
% appendix
{
\onecolumn
\appendix
\section*{Appendix}

From the boundary conditions, Eqs.
\eqref{eq:bCond1}--\eqref{eq:bCond4}, the dispersion relation, Eq.
\eqref{eq:DispRel}, is expressed through $\tilde D_1= D_1/D_0$ and
$\tilde D_2=D_2/D_0$, where

\begin{align}
D_0=& {{\mr^2}}\,{\nma}\,{\npa}\,
  {\sma}\,{\spa}\,{\left( {{\ub}}^2 - {{\vb}}^2 \right)}^2,
\\
D_1=& F_0 + F_1(\atz) + F_2(\atz)^2,
\\
D_2=& G_0 + G_1 (\atz) + G_2 (\atz)^2
   + G_3 (\atz)^3 + G_4 (\atz)^4.
\end{align}%
Here, $\mr=m_N/m_S$, $F_0$ through $F_2$ and $G_0$ through
$G_4$ are given by

%%%%%%%%%%%%%%%%%%%%%%%%%%%%%%%%%%%%%%%%%%%%%%%%%%%%%%%%%
%%%%%%%%%%%%%%%%%%%%%%%%%%%%%%%%%%%%%%%%%%%%%%%%%%%%%%%%%
{\small %
\begin{align}
F_0=&
{\mr}\,
\left( {{\ub}}^2 -{{\vb}}^2 \right) \,
\Big\{ %%%
    \big[
        {\npa}\,{\spa}\,{\snm}\,{\ssm}
        \left(
            {{\nma{}^2}} +{{\mr^2}}\,{{\sma{}^2}}
        \right)
        +
        {\nma}\,{\sma}\,{\snp}\,{\ssp}
        \left(
            {{\npa{}^2}} + {{\mr^2}}\,{{\spa{}^2}}
        \right)
    \big]
    \,{{\ub}}^2
%-
\notag
\\
&\qquad\qquad\qquad\quad -
    \big[
        {\nma}\,{\spa}\,{\snp}\,{\ssm}
        \left(
            {{\npa{}^2}} + {{\mr^2}}\,{{\sma{}^2}}
        \right)
        +
        {\npa}\,{\sma}\,{\snm}\,{\ssp}
        \left(
        {{\nma{}^2}} + {{\mr^2}}\,{{\spa{}^2}}
        \right)
    \big]
    \,{{\vb}}^2
%+
\notag
\\
&\qquad\qquad\qquad\quad +
    2\,{\mr}\,{\nma}\,{\npa}\,{\sma}\,{\spa}\,
    \big[
        \vb^2
        \left(
            \cnp\csm + \cnm\csp
        \right)
        -
        \ub^2
        \left(
            \cnp\csp + \cnm\csm
        \right)
    \big]
\Big\},  %%%
\end{align}%
%%%%%%%%%%%%%%%%%%%%%
\begin{align}
F_1 =&
-2\,{{\mr^2}}\,
\left(
    {{\ub}}^2 - {{\vb}}^2
\right)
\,
\Big\{
    {\nma}\,{\npa}\,
    \big[
        \spa \, \ssm
        \left(
            \cnm \, \ub^2 - \cnp \, \vb^2
        \right)
        +
        \sma \, \ssp
        \left(
            \cnp \, \ub^2 - \cnm \, \vb^2
        \right)
    \big]
%+
\notag
\\
&\qquad\qquad\qquad\qquad\qquad+
    {\mr}\,{\sma}\,{\spa}\,
    \big[
        \nma \, \snp
        \left(
            \csp \, \ub^2 - \csm \, \vb^2
        \right)
        +
        \npa \, \snm
        \left(
            \csm \, \ub^2 - \csp \, \vb^2
        \right)
    \big]
\Big\},
\end{align}%
%%%%%%%%%%%%%%
\begin{align}
F_2 =&
-
{{\mr^3}}\,
\left(
    {{\ub}}^2 - {{\vb}}^2
\right)
\,
\big[
    {\nma}\,{\snp}\,
    \left(
        {\sma}\,{\ssp}\,{{\ub}}^2
        -
        {\spa}\,{\ssm}\, {{\vb}}^2
    \right)
    +
    {\npa}\,{\snm}\,
    \left(
        {\spa}\,{\ssm}\,{{\ub}}^2
        -
        {\sma}\,{\ssp}\,{{\vb}}^2
    \right)
\big],
\end{align}%
}% scriptsize
%%%%%%%%%%%%%%%%%%%%%%%%%%%%%%%%%%%%%%%%%%%%%%%%%%%%%%%%%
%%%%%%%%%%%%%%%%%%%%%%%%%%%%%%%%%%%%%%%%%%%%%%%%%%%%%%%%%
and
%%%%%%%%%%%%%%%%%%%%%%%%%%%%%%%%%%%%%%%%%%%%%%%%%%%%%%%%%
%%%%%%%%%%%%%%%%%%%%%%%%%%%%%%%%%%%%%%%%%%%%%%%%%%%%%%%%%
{\small %
\begin{align}
G_0=&
{\snm}\,{\snp}\,{\ssm}\,{\ssp}\,
\left(
    {{\nma{}^2}}\,{{\npa{}^2}}
    +
    {{\mr^4}}\,{{\sma{}^2}}\,{{\spa{}^2}}
\right)
{\left(
    {{\ub}}^2 - {{\vb}}^2
\right)}^2
%\times
\notag
\\
%%%%%%%
&
-
2\,
\left(
    {{\ub}}^2 - {{\vb}}^2
\right)
\,
%\times
\Big\{
    {\mr}\,{\nma}\,{\npa}\,
    \big[
        {\npa}\,{\cnm}\,{\snp}\,
        \left(
            {\sma}\,{\csm}\,{\ssp}\,{{\ub}}^2
            -
            {\spa}\,{\csp}\,{\ssm}\,{{\vb}}^2
        \right)
        +
        {\nma}\,{\cnp}\, {\snm}\,
        \left(
            {\spa}\,{\csp}\,{\ssm}\,{{\ub}}^2
            -
            {\sma}\,{\csm}\,{\ssp}\,{{\vb}}^2
        \right)
    \big]
    %+
    \notag
    \\
    %%%%%%
    &\qquad\qquad\qquad\qquad+
    {{\mr^3}}\,{\sma}\,{\spa}\,
    \big[
        {\nma}\,{\cnm}\,{\snp}\,
        \left(
            {\spa}\,{\csm}\,{\ssp}\,{{\ub}}^2
            -
            {\sma}\,{\csp}\,{\ssm}\,{{\vb}}^2
        \right)
        +
        {\npa}\,{\cnp}\,{\snm}\,
        \left(
            {\sma}\,{\csp}\,{\ssm}\,{{\ub}}^2
            -
            {\spa}\,{\csm}\,{\ssp}\,{{\vb}}^2
        \right)
    \big]
\Big\}
%+
\notag
\\
%%%%%%%%
&+
{{\mr^2}}\,
{{\Big(}}
    2\,{\nma}\,{\npa}\,
    \big\{
        {\ssm}\,{\ssp}\,{{\ub}}^2\,{{\vb}}^2
        \left(
            {\cnm}\,{\cnp}-1
        \right)
        \,
        \left(
            {{\sma{}^2}} + {{\spa{}^2}}
        \right)
        %+
        \notag
        \\
        %%%%%%%%%%%
        &\qquad\qquad\qquad\qquad+
        {\sma}\,{\spa}\,
        \big[
            \left(
                1 + 2\,{\cnm}\,{\cnp}\,{\csm}\, {\csp}
            \right)
            \,
            \left(
                {{\ub}}^4 + {{\vb}}^4
            \right)
            -2\,{{\ub}}^2\,{{\vb}}^2
            \left(
                {\csm}\,{\csp}
                +
                {\cnm}\,{\cnp}
                +
                {\cnm}\,{\cnp}\,{\csm}\,{\csp}
            \right)
        \big]
    \big\}
    %+
    \notag
    \\
    %%%%%%%%%%%%%%
    &\qquad\qquad+
    {\snm}\,{\snp}\,
    \big\{
        2 \ksm \, \ksp \, \ub^2 \, \vb^2 \,
        \left(\csm \,\csp-1\right)
        \left(\knm{}^2+\knp{}^2 \right)
        %+
        \notag
        \\
        %%%%%%%%%%
        &\qquad\qquad\qquad\qquad+
        \ssm \, \ssp
        \big[
            \ub^4
            \left(
            \knm{}^2 \, \ksp{}^2 + \knp{}^2 \, \ksm{}^2
            \right)
            +
            \vb^4
            \left(
            \knm{}^2 \, \ksm{}^2 + \knp{}^2 \, \ksp{}^2
            \right)
        \big]
    \big\}
{{\Big)}},
\end{align}%
}% scriptsize
%%%%%%%%%%%%%%%%%%%%%%%%%%%%%%%%%%%%%%%%%%%%%%%%%%%%%%%%%
%%%%%%%%%%%%%%%%%%%%%%%%%%%%%%%%%%%%%%%%%%%%%%%%%%%%%%%%%
{\small %
\begin{align}
G_1 = &
-2\,{\mr}\,
\left(
    {{\ub}}^2 - {{\vb}}^2
\right)
\notag
\\
%%%%%%%%%%
& \times
{{\Big[}}
    {\snp}\,
    {{\Big(}}
        {\mr}\,{\snm}\,
        \big\{
            {\csp}\,{\spa}\,{\ssm}\,
            \big[
                {{\nma{}^2}}\,{{\ub}}^2
                -
                {{\npa{}^2}}\,{{\vb}}^2
                +
                {{\mr^2}}\,{{\sma{}^2}}\,
                \left(
                    {{\ub}}^2 - {{\vb}}^2
                \right)
            \big]
            %+
            \notag
            \\
            %%%%%%%%%%
            &\qquad\qquad\qquad\qquad +
            {\csm}\,{\sma}\,{\ssp}\,
            \big[
                {{\npa{}^2}}\,{{\ub}}^2
                -
                {{\nma{}^2}}\,{{\vb}}^2
                +
                {{\mr^2}}\,{{\spa{}^2}}\,
                \left(
                    {{\ub}}^2 - {{\vb}}^2
                \right)
            \big]
        \big\}
        %+
        \notag
        \\
        %%%%%%%%%
        &\qquad\qquad +
        {\cnm}\,{\nma}\,
        \big\{
            -2\,{\csm}\,{\csp}\,{{\mr^2}}\,{\sma}\, {\spa}\,
            \left(
                {{\ub}}^2 -  {{\vb}}^2
            \right)
            +
            {\ssm}\,{\ssp}\,
            \big[
                {{\npa{}^2}}\,
                \left(
                    {{\ub}}^2 - {{\vb}}^2
                \right)
                +
                {{\mr^2}}\,
                \left(
                    {{\spa{}^2}}\, {{\ub}}^2
                    -
                    {{\sma{}^2}}\,{{\vb}}^2
                \right)
            \big]
        \big\}
    {{\Big)}}
    %+
    \notag
    \\
    %%%%%%%
    &\qquad +
    {\cnp}\,{\npa}\,
    {{\Big(}}
        -2\,{\csm}\,{\mr}\,{\sma}\,
        \left(
            {\csp}\,{\mr}\,{\snm}\,{\spa}
            +
            {\cnm}\,{\nma}\,{\ssp}
        \right)
        \,
        \left(
            {{\ub}}^2 -{{\vb}}^2
        \right)
        %+
        \notag
        \\
        %%%%%%%%
        &\qquad\qquad\qquad\quad\! +
        {\ssm}\,
        \big\{
            -2\,{\cnm}\,{\csp}\, {\mr}\,{\nma}\, {\spa}\,
            \left(
                {{\ub}}^2 - {{\vb}}^2
            \right)
            +
            {\snm}\,{\ssp}\,
            \big[
                {{\nma{}^2}}\,
                \left(
                    {{\ub}}^2 - {{\vb}}^2
                \right)
                +
                {{\mr^2}}\,
                \left(
                    {{\sma{}^2}}\,{{\ub}}^2 - {{\spa{}^2}}\,{{\vb}}^2
                \right)
            \big]
        \big\}
    {{\Big)}}
{{\Big]}},
\end{align}%
}% scriptsize
%%%%%%%%%%%%%%%%%%%%%%%%%%%%%%%%%%%%%%%%%%%%%%%%%%%%%%%%%
%%%%%%%%%%%%%%%%%%%%%%%%%%%%%%%%%%%%%%%%%%%%%%%%%%%%%%%%%
{ \small
\begin{align}
G_2 = &
-{{\mr^2}}\,
\left(
    {{\ub}}^2 - {{\vb}}^2
\right)
\notag
\\
%%%%%%%%%%
& \times
{{\Big(}}
    2\,{\csm}\,{\mr}\,{\sma}\,
    \big\{
        2\,{\csp}\,{\mr}\,{\snm}\,{\snp}\,{\spa}\,
        \left(
            -{{\ub}}^2 + {{\vb}}^2
        \right)
        +
%        \notag
%        \\
        %%%%%%%
%        &+
        {\ssp}\,
        \big[
            -{\cnm}\,{\nma}\,{\snp}\,
            \left(
                {{\ub}}^2 - 2\,{{\vb}}^2
            \right)
            +
            {\cnp}\,{\npa}\,{\snm}\,
            \left(
                -2\,{{\ub}}^2 + {{\vb}}^2
            \right)
        \big]
    \big\}
    %+
    \notag
    \\
    %%%%%%%%%%%
    &\qquad +
    {\ssm}\,
    {{\big(}}
        -2\,{\cnp}\,{\npa}\,
        \big[
            {\csp}\,{\mr}\,{\snm}\,{\spa}\,
            \left(
                {{\ub}}^2 - 2\,{{\vb}}^2
            \right)
            +
            2\,{\cnm}\,{\nma}\, {\ssp}\,
            \left(
                {{\ub}}^2 - {{\vb}}^2
            \right)
        \big]
        %+
        \notag
        \\
        %%%%%%%%%%%
        &\qquad\qquad\quad +
        {\snp}\,
        \big\{
            {\snm}\,
            \big[
                {{\nma{}^2}}
                +
                {{\npa{}^2}}
                +
                {{\mr^2}}\,
                \left(
                    {{\sma{}^2}} + {{\spa{}^2}}
                \right)
            \big]
            \, {\ssp}\,
            \left(
                {{\ub}}^2 - {{\vb}}^2
            \right)
            +
            2\,{\cnm}\,{\csp}\,{\mr}\,{\nma}\,{\spa}\,
            \left(
                -2\,{{\ub}}^2 + {{\vb}}^2
            \right)
        \big\}
    {{\big)}}
{{\Big)}},
\end{align}%
}% scriptsize
%%%%%%%%%%%%%%%%%%%%%%%%%%%%%%%%%%%%%%%%%%%%%%%%%%%%%%%%%
%%%%%%%%%%%%%%%%%%%%%%%%%%%%%%%%%%%%%%%%%%%%%%%%%%%%%%%%%
{ \small
\begin{align}
G_3 =
2\,{{\mr^3}}\,
{\left(
    {{\ub}}^2 - {{\vb}}^2
\right)}^2
\,
\big\{
    {\mr}\,{\spa}\,{\snm}\,{\snp}\,{\ssm}\,{\csp}
    +
    {\ssp}\,
    \big[
        {\nma}\,{\cnm}\,{\snp}\,{\ssm}
        +
        {\snm}\,
        \left(
            {\mr}\,{\sma}\,{\csm}\,{\snp}
            +
            {\npa}\,{\cnp}\,{\ssm}
        \right)
    \big]
\big\},
\end{align}%
}% scriptsize
%%%%%%%%%%%%%%%%%%%%%%%%%%%%%%%%%%%%%%%%%%%%%%%%%%%%%%%%%
%%%%%%%%%%%%%%%%%%%%%%%%%%%%%%%%%%%%%%%%%%%%%%%%%%%%%%%%%
{ \small
\begin{align}
G_4 =
{{\mr^4}}\,{\snm}\,{\snp}\,{\ssm}\,{\ssp}\,
{\left(
{{\ub}}^2 - {{\vb}}^2
\right)}^2,
\end{align}%
}% scriptsize
%%%%%%%%%%%%%%%%%%%%%%%%%%%%%%%%%%%%%%%%%%%%%%%%%%%%%%%%%
%%%%%%%%%%%%%%%%%%%%%%%%%%%%%%%%%%%%%%%%%%%%%%%%%%%%%%%%%
where
$s^\pm_N\equiv\sin(k^\pm_N \dn)$,
$c^\pm_N\equiv\cos(k^\pm_N\dn)$,
$s^\pm_S\equiv\sin(k^\pm_S \ds)$, and
$c^\pm_S\equiv\cos(k^\pm_S \ds)$.
For $Z=0$ and $\mr=1$ expressions for $\tilde D_1$ and
$\tilde D_2$ reduce to the results given in Ref.
\cite{art:Kummel94}.

%\end{widetext}
\twocolumn
} %appendix
%%%%%%%%%%%%%%%%%%%%%%%%%%%%%%%%%%%%%%%%%%%%%%%%%%%%%%%%%%
%%%%%%%%%%%%%%%%%%%%%%%%%%%%%%%%%%%%%%%%%%%%%%%%%%%%%%%%%%

\end{document}